\documentclass[12pt]{article}
\usepackage[tmargin=1in,bmargin=1in,lmargin=1.25in,rmargin=1.25in]{geometry}

\usepackage{setspace}
\usepackage{csquotes}
\usepackage{amsmath}
\usepackage{amsfonts}
\usepackage{stmaryrd}
\usepackage{dsfont}
\usepackage{graphicx}
\usepackage[title]{appendix}
\usepackage[font=small,labelfont=bf]{caption}
\usepackage{lipsum}
\usepackage{mwe}
\graphicspath{ {images/} }

\title{Can we ``effectivize'' spacetime?}
\date{Lu Chen}
\author{Published in \textit{Studies in History and Philosophy of Science}}
\begin{document}
	\maketitle
	
\paragraph{Abstract} According to \textit{effective realism}, scientific theories give us knowledge about the unobservable world, but not at the fundamental level. This view is supported by the well-received \textit{effective-field-theory} (EFT) approach to high energy physics, according to which even our most successful physical theories are only applicable up to a certain energy scale and expected to break down beyond that. In this paper, I advance new challenges for effective realism and the EFT approach. I argue that \textit{effective quantum gravity} (EQG) does not give us a realistic theory of spacetime even within its scope of validity. This also exposes a general interpretative dilemma faced by all EFTs concerning their indispensable references to classical spacetime beyond their scope of validity. 


\paragraph{}
	
What is the world like at the fundamental level according to our best scientific theories? This is a question often asked by philosophers, especially those who endorse scientific realism. However, we have good reasons to believe that this question relies on a faulty assumption about science and should be abandoned. I do not mean the reasons given by the traditional foes of realism---in particular, those who argue that science does not aim at providing us knowledge beyond the empirical realm. Rather, I mean the novel challenges presented by a contemporary approach to high energy physics, according to which even \textit{quantum field theories} (QFTs)---which are arguably our  ``most powerful, beautiful and effective theoretical edifice ever constructed in the physical sciences'' (Duncan 2012, iv)---are considered as only applicable to the presently experimentally accessible energy levels and are expected to break down beyond that level. Such theories are called the \textit{effective field theories} (EFTs) and the approach, \textit{the EFT approach}.\footnote{In this paper, I will almost exclusively focus on high energy physics (with occasional exceptions) where the techniques of QFTs are relevant, as opposed to classical physics and other sciences, since high energy physics is a crucial frontline in our scientific progress towards the unknown and arguably at the center of the debate about scientific realism. However, the discussion of this paper may be carried over to other theories. For example, the idea that a theory should be trusted only within a limited scope can be generalized to classical physics and more, although I will not explore it in-depth in this paper. } According to this approach, we may never be able to plausibly assert that we have found the fundamental theory. This approach is well received by the physics community, and appropriately so, as I will explain in Section 1.  What is the implication of this approach for scientific realism and our worldview? 

At first, it may sound like a triumph of anti-realism: we are only licensed to believe in the empirical predictions of our best theories because they are merely \textit{effective}. But this is not so. EFTs make a wide range of theoretical statements about entities that are standardly considered unobservable such as electrons, neutrons, and quarks. The fact that EFTs are expected to break down beyond a certain scale does not automatically entail that we should not believe in the (approximate) truth  of their theoretical statements within the scope of those theories. Indeed, Williams (2019) and Fraser (2018) among others argue that the EFT approach to physics can be reconciled with realism, which yields an attractive novel implementation of selective realism---call it \textit{effective realism}---according to which only the ontology within the validity scope has realistic significance. Effective realism provides principled ways to single out entities worthy of ontological commitment within the domain of a scientific theory (Section 2). 

Nonetheless, I will point out some important challenges facing the general strategy of effective realism. First, I shall apply this strategy to \textit{effective quantum gravity} (EQG). In the classical picture, spacetime metrics are well defined for arbitrarily small regions, and larger distances are obtained by integrating over local metrics. It has been long realized that this classical picture is inaccurate for the small realm where quantum physics takes over---metrics too are governed by quantum dynamics. In the philosophical literature, we have not talked about this picture much mainly because it is commonly assumed  that we know very little about spacetime beyond general relativity due to the incompatibility between general relativity and quantum physics. This assumption needs to be qualified because the incompatibility in question is true only if we want the unified theory to be a fundamental theory---a theory amenable to empirical tests however small scales we probe---a desire we should forsake according to the EFT approach. At currently experimentally accessible scales, we do have a physically well-defined unification of general relativity and quantum theory, namely EQG (Section 3). 

Can we extract a realistic theory of spacetime metrics at accessibly small scales from EQG under the strategy of effective realism? Although I hope this is the case, my conclusion is nevertheless negative (Section 3). The problem has to do with the EQG's reliance on a classical background metric, which is defined to arbitrary accuracy. If we follow the standard prescription that we use in interpreting other EFTs, we will commit to an undesirable dualism between the graviton field and the background metric.  Moreover, we face a dilemma in whether to take the background metric realistically because it is defined beyond the validity scope and yet plays an indispensable role in the theory.   

The second difficulty is not limited to the interpretation of EQG but present to all EFTs, since they all involve references to the classical background metric (Section 4). It may be tempting  to exclude spacetime from the ontology of the EFTs that are \textit{matter} theories, which are not obliged to offer an exhaustive description of reality and in particular of spacetime. But the references to background spacetime play an important explanatory role in those theories and therefore should not be dispensed with in our interpretation. I do not intend this to be a devastating argument against effective realism. The moral is that we should address these important challenges to effective realism, which is otherwise attractive and even a necessary adaptation of realism to the EFT approach to physics. To the extent that scientific realism is attractive, these are also challenges for the very EFT approach.

As often is the case, technicality is a double-edged sword---it helps the discussion to be more scientifically accurate and non-superficial, but it can also unnecessarily distract or intimidate the general reader away from the philosophical gist.  In the hope of dulling its bad edge, I would like to emphasize that many technical details (e.g., Section 1.1, 1.2, 3.1) are skimmable if the reader can accept the main technical claims  relevant  to the philosophical discussion.

	\section{A Short Introduction to EFTs}
	
	The techniques of \textit{renormalization} are at the core of the framework of QFTs. It was once perceived as a necessary criterion for physical theories that they are \textit{perturbatively renormalizable}.  In this section, I will explain what this means, why our attitude towards it has changed, and how this change leads to the EFT approach to physics. I will argue that our best scientific theories should be considered EFTs---valid up to a certain scale.  (For detailed expositions on EFTs, see for example Costello 2011, Butterfield and Bouatta 2014, Crowther and Linnemann 2017, and Williams 2021)

	\paragraph{1.1 What is perturbative renormalizability?}
	
	Consider a scalar field $\phi$ with mass $m$ and an interaction coupling constant $\lambda$, which is needed for describing how the field interacts with itself.\footnote{The underlying theory is called \textit{ the $\phi^4$ theory}, which is often used as the simplest (toy) example of a renormalizable theory. (See Peskin and Schroeder 1995)} In the framework of QFT, we can calculate the probabilistic correlation between certain values of $\phi$ at four spacetime points, which is called \textit{4-point function}. This correlation is closely related to  the amplitude (which gives us the probability) of two particles  (as excitations of the field) scattering with certain energies and angles and producing another two particles with certain energies
	and angles.\footnote{The correlation function allows scattering amplitudes to be calculated through the LSZ-reduction formula. Henceforth I will not emphasize the distinction between these two closely related quantities.} The value of the correlation function is given by integrating over all possible configurations of $\phi$ with the given values at the four spacetime points $\phi(x_1),\phi(x_2),\phi(x_3),\phi(x_4)$. The integrand is given by the Lagrangian $\mathcal{L}$ of the field, which encodes all its dynamical information. We can write it out like this:	$$ \mathcal{A}=\int D\phi \phi(x_1)\phi(x_2)\phi(x_3)\phi(x_4) e^{i\int d^4x \mathcal{L}(m,\lambda)} $$
($D\phi$ signifies integrating over all field configurations; $d^4 x$ signifies integrating over all spacetime points.) Unfortunately, when we try to calculate $\mathcal{A}$ by expanding it into a Taylor series ordered by the magnitude of $\lambda$, we see that most terms are divergent, and adding them up is not mathematically well-defined. More concretely, when calculating the amplitude of the aforementioned scattering, we need to integrate over all the possible interactions happening in the process, such as the two particles exchanging ``photons'' any number of times.  But those virtue ``photons" can have arbitrarily high momenta and integrating over them leads to divergence. 

Here's where the technique of renormalization comes to the rescue. The first step is called \textit{regularization}. There are many ways to regularize, among which a common one is to impose a momentum cut-off $\Lambda$.
 The idea is that if we ignore all the configurations of $\phi$ that involve momenta above that cut-off point, then the integral in question becomes finite.
 $$	\int^\Lambda D\phi ... = \textrm{finite}.$$
 (We can also think of it in terms of length, which is inversely related to momentum: the regularization amounts to ignoring all fluctuations of the field occuring within the cut-off length.) After the regularization, the value of the integral would appear to depend on the cut-off points (the larger $\Lambda$ is, the larger the integral is). But we cannot allow this because the value of the integral is an objective physical quantity while the cut-off point is arbitrarily chosen. The second step of renormalization precisely addresses this problem: we calculate how the parameters of the field $m,\lambda$ vary with the cut-off point $\Lambda$ so that the value of the integral $\mathcal{A}$ converges to a finite value as as $\Lambda$ tends to infinity.
$$	\mathcal{A}(m(\Lambda),\lambda(\Lambda),...)_{\Lambda\to \infty} = \textrm{finite}.$$
 If we can solve this equation by specifying $m,\lambda$ as functions of $\Lambda$, then the theory is \textit{perturbatively renormalizable}.\footnote{Strictly speaking, this definition merely refers to  the old-school conception of renormalization as opposed to the new-school Wilsonian understanding of it, which will be discussed soon in Section 1.2. Like others in the literature, I use the label ``perturbative renormalizability'' because perturbation theory is typically used in this old-school context.

When calculating $m,\lambda$ as functions of $\Lambda$, we also need to take the empirical constraints into account. I will talk more about this later.} If necessary, we can introduce new parameters to $\mathcal{A}$ that also depend on the cut-off point. If introducing finitely many such parameters can solve such an equation, then we also say that the theory is perturbatively renormalizable.\footnote{In this particular case, namely the $\phi^4$ theory, no additional parameter is needed (see Peskin and Schroeder 1995, 324-5).} Otherwise, it is perturbatively non-renormalizable. Non-renormalizability is undesirable because if $\mathcal{A}$ contains infinitely many parameters, then no matter how many observations we make, we will never be able to obtain or verify values of those infinitely many parameters. In other words, if a theory is defined by infinitely many parameters, it is nonverifiable with finitely many experiments. As such, renormalizability was considered an important criterion for whether a theory is physical (that is, finitely verifiable). 

But after the second step, the mass $m$ and the coupling constant $\lambda$ become dependent on the cut-off point, and may even go to infinity as the cut-off tends to infinity. One may wonder: aren't they physical quantities? But physical quantities should not depend on arbitrary choices of cut-off points nor be infinite.\footnote{Many people accept actual infinities and do not regard them as a problem: for instance, a line segment is composed of infinitely many points. But actual infinities become problematic if we claim them to be measurable quantities, since it is certain unusual to have the ``Infinity'' reading on our measurable devices. Fortunately, bare constants are not measuring quantities. For renormalizable theories, physical constants do have finite values.} Indeed, it is best not to think of these constants as physical quantities but theoretical posits that are partly artifacts of our regularization scheme---we call them \textit{bare constants}. In contrast, the \textit{physical constants} that we measure in experiments are functions of the bare constants, the cutoff point, and the energy levels of our experiments.\footnote{Although physical constants are not directly measurable---what we measure more directly are the values of correlation functions or the corresponding amplitudes---it's not a great stretch to consider them measurable.}  So, once we know how the bare constants vary with the cut-off point, we can calculate the values of the physical constants at the energy levels of our experiments and compare it with our experimental results. Or rather, given the values of physical constants we measure in experiments and the energy levels of those experiments, which are the empirical constraints, we can calculate how bare constants vary with the cut-off points so that the constraints are satisfied. Magically enough, the physical constants are indeed independent from the cut-off point (their dependence on the bare constants and the cutoff “cancel off”). They only depend on the energy level of our experiments: for example, the measured charge of an electron depends on the energy we use to probe the electron.

\paragraph{1.2 Why is perturbative renormalizability now considered less important?}
The dominant attitude towards renormalizability has changed under the rise of the \textit{renormalization group} (RG) method  since the 1980s.  Instead of the bare coupling constants and the cut-off points, RG focuses on how \textit{physical} coupling constants vary with the energy scale of the experiment (or inversely, the length scale). For the scalar field $\phi$, if we know how the bare constants vary with the cut-off points, we can also straightforwardly obtain how the values of the physical constants vary with the experimental energy scale $\mu$ with the cut-off point fixed, given that  the functions of the physical constants on the bare constants, the cut-off points and the energy scales are known. For concreteness, the equation for the physical coupling constant $\lambda_p$ as a function of energy scale $\mu$ can be written as follows ($c$ is a real number constant; $\mathcal{O}$ abbreviates less important terms):
$$\mu\frac{d}{d\mu}\lambda_p(\mu)=c\lambda_p(\mu)^2+\mathcal{O}$$
This equation is called the \textit{renormalization group (RG) equation} for $\lambda_p$ (see Williams 2021). More generally, for any theory with physical constants $g_1,g_2...,g_n$, the RG equation for a given $g_i$ can involve all these constants:
$$	\mu\frac{d}{d\mu}g_i(\mu)=\beta_i(g_1,...,g_n)$$
 If we picture the values of $n$ physical coupling constants at a given energy scale as a point in an $n$-dimensional abstract space, then their varying with the energy scale can be pictured as the point moving around in the space as we adjust the energy scale. This ``motion picture'' is called \textit{running coupling constants} or \textit{renormalization group flow}. If the coupling constants have finite limits as the energy scale tends to infinity, then we say the theory is \textit{non-perturbatively renormalizable}.\footnote{Strictly speaking, the definition mainly refers to the new-school conception of renormalizability (see also Footnote 4). I use the label ``non-perturbative'' because the technique of perturbation theory is less important in this context. However, perturbation theory can still be used. } Importantly, it turns out that some perturbatively non-renormalizable theories are non-perturbatively renormalizable (for example, see Braun et al.\@ 2011). 

An important insight from the RG method is that non-renormalizability is not necessarily a problem. Recall that the problem of a (perturbatively) non-renormalizable theory lies in its nonverifiability (with finitely many experiments). However, the non-renormalizable terms in the Lagrangian of a  non-renormalizable theory that seem so troublesome need not be a problem if we exclusively focus on the theory at sufficiently low energy scales, among which are currently experimentally accessible levels (Figure 1). 
\begin{figure}[h]
	\centering
	\includegraphics[width=1\linewidth]{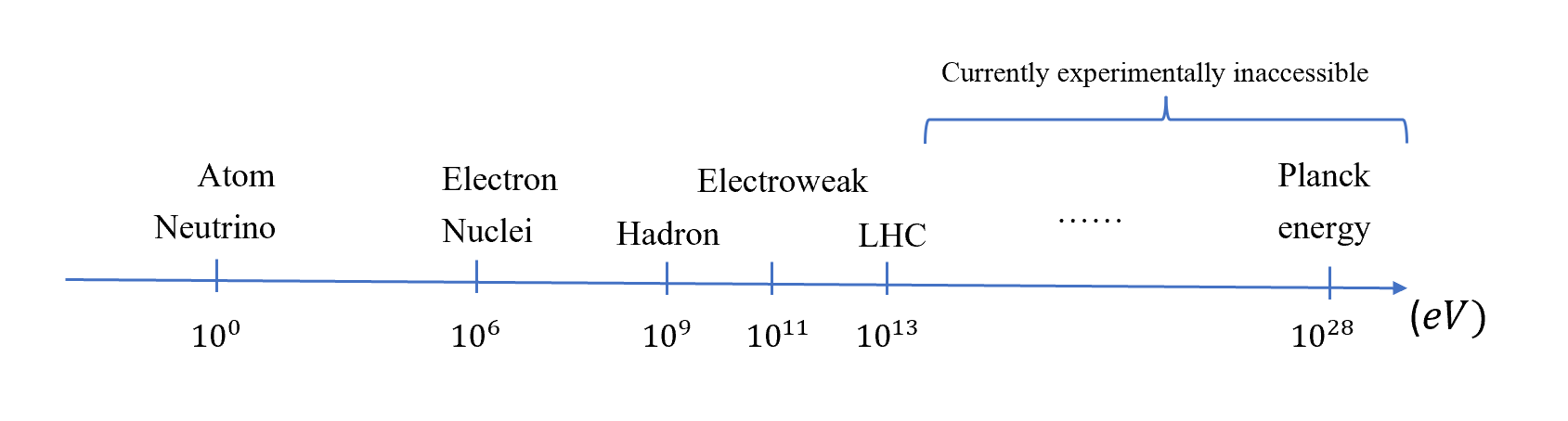}
	\caption[Energy scales in physics]{Energy scales in physics (Hertel and Schulz 2015)}
	\label{fig:energy}
\end{figure}

\noindent  The reason is roughly that, because those non-renormalizable terms are multiplied by physical coupling constants, which get smaller as the energy level goes lower, they become irrelevant at the low energy level (see Wilson 1979). In other words, if we restrict the scope of the theory to the low energy range, then we can ignore the non-renormalizable terms. As a result, we can extract empirical predictions from the theory at the low energy level just like from a renormalizable theory. 

\paragraph{1.3 What is the EFT approach to physics?}
The RG method naturally leads to the EFT approach according to which (high energy) physical theories are EFTs---valid only up to a scale. Suppose we only care about the energy levels below $\mu_0$ that we currently have access to. Then, we can impose a momentum cut-off point $\Lambda_0$ suitably beyond $\mu_0$ ($\Lambda_0\gg\mu_0$) which allows us to ignore all interactions that involve energies beyond $\Lambda_0$. For example, if we try to model the scattering of protons at the energy level of about 1 GeV, then we can impose a cutoff point suitably larger than 1 GeV but smaller than the energy level of quarks and gluons (about $10^{15}$ GeV). This allows us to ignore the quarks and gluons and in general higher-energy degrees of freedom so that we can optimize parameters to model the system efficiently at the low-energy range. 
(This is very intuitive also for non-QFTs: for instance, we know that in describing the heat transfer of a macro-system, it is sufficient to use macro-parameters like temperature and conductivity of the material---the micro-parameters like the Brownian motions of the molecules are largely irrelevant.) The resulting theory is valid only up to an energy level suitably below the cut-off point. If we conduct an experiment at the energy level near or beyond the cut-off point, then we can expect the theory's predictions become very inaccurate, as the finer details blurred out by the cut-off point become relevant.

\paragraph{1.4 Why should QFTs be considered EFTs?}
The physics community has largely shifted from the attitude that renormalization is a necessary condition for a physical theory to the attitude that even the current QFTs, which are renormalizable and arguably most impressive achievements in physics so far, are just EFTs. As I will argue, this shift is warranted because we should exercise high energy humility. 

First, I want to reject a consideration mentioned in Butterfield and Bouatta (2014, 19-20) among others: we should consider QFTs as EFTs because otherwise it would be ``a piece of great good fortune'' that QFTs happen to be perturbatively renormalizable. The idea is that non-renormalizable theories (e.g., the Fermi theory of the weak interaction) are perfectly conceivable. According to Dyson's power-counting method, a theory is renormalizable only if the Lagrangian does not contain coupling constants with a negative dimension of mass, but we can conceive many theories that have such constants (e.g., see Peskin and Schroeder 1995: 315-323). Taking QFTs as EFTs can explain away this lucky coincidence, because within the low energy level all non-renormalizable terms are suppressed by small coefficients associated with the low energy level and therefore do not contribute to the result. 

However, this argument seems incompatible with the wide acceptance of renormalizability  as a necessary condition for a fundamental or final theory.\footnote{More precisely, nonperturbative renormalizability is now considered a criterion for fundamental theories (see Crowther and Linnemann 2017). But which notion of renormalizability is used here does not matter much, as neither needs to be explained away. } As long as we do not want to necessarily avoid the lucky coincidence of having a final theory,  renormalizability isn't exactly a coincidence that has to be explained away. Having conceivable and coherent alternatives does not necessarily mean our physically well-behaved theory requires a miracle---it shouldn't be too much to ask of our physical reality to be well-behaved! I think a better rationale for taking QFTs as EFTs is that we should not take QFTs as our final theories, not that their renormalizability would otherwise be a miracle.\footnote{This may be what Butterfield and Bouatta (2014, 19-20) have in mind. If that's the case, I hope to make the point more explicit and ward off a potential confusion.}

 As we all know, atoms, once considered basic, are composed of neutrons, protons and electrons, and neutrons and protons in turn are composed of quarks. There is no reason to think that quarks, which are considered fundamental particles in the standard model of QFTs, are not composed of smaller entities.  We do not know whether new particles and interactions will appear at a higher energy level, so the most responsible epistemic attitude is to exercise high energy humility---that is, to remain agnostic about what physics is like at the high energy level. Indeed, recent experimental results indicate a decent chance that we have found new degrees of freedom beyond the standard model: the B-meson decays observed in LHCb seem to deviate from the predictions of the standard model, and some researchers have cautiously suggested the phenomena to be explained by new particles called leptonquarks and charged singlets (Marzocca and Trifinopoulos 2021).

More technically, the fact that QFTs are empirically successful at the low energy level that we have access to is no indication of how they will apply to higher-energy levels because the low energy theories are ``decoupled'' from their higher energy counterparts: two very similar EFTs that are practically indistinguishable at their validity range can have very different underlying higher-energy theories---and vice versa. In terms of RG flows, closely grouped points that represent different theories at a certain energy level can flow in diverging ways not only as the energy level goes higher but also as it goes lower. For example, theories that belong to ``a universality class'' converge to a fixed point in their RG flow (see Butterfield and Bouatta 2014). Note, however, that low energy theories are still in principle reducible to their higher-energy counterparts, and although such a reduction is often extremely difficult in practice, it has been done in some cases (see Petreczky et al.\@ 2018). So, the decoupling in question is not complete. But this does not matter for my argument: our empirical measures are not fine enough to differentiate between similar low-energy theories with different higher-energy versions. In particular, we do not have enough empirical evidence to differentiate between QFTs that apply to all energy levels and their slightly modified counterparts QFTs* that are similar in the low energy level but run in diverging courses as the energy level increases. 

Not only the QFTs at currently experimentally accessible levels are decoupled from higher-energy theories, but the whole framework of QFT itself is also decoupled from a possible higher-energy alternative. Weinberg (1995, 1999) has shown that any quantum theory satisfying Lorentz invariance at the low energy level and one other plausible assumption called ``cluster decomposition" must be a QFT at the low energy level. So the fact that the framework of QFTs is empirically successful at our experimentally accessible energy level is not strong evidence for its applicability to higher-energy levels. Indeed, the sustained difficulty in unifying quantum theory with general relativity at higher energy levels (e.g., the Planck level) calls for a new framework at those levels.

\section{Effective Realism}

If we take our (high energy) physical theories as merely EFTs---valid up to a certain energy scale, how does this affect the outlook of scientific realism, the view that scientific theories give us (or aim at) knowledge beyond empirical predictions?  Contrary to what one might initially think, reconciling the EFT approach with scientific realism is not only possible but also leads to an attractive version of realism, as argued by  Williams (2019) and Fraser (2018) among others.\footnote{The versions of realism these authors formulate are slightly different. In this section, my presentation mainly follows Williams' version. Ruetsche (2018) has also proposed a version of effective realism (which she calls ``renormalization group realism'') in the framework of EFTs. It is worth mentioning that she has also expressed doubt about whether effective realism can meet the skeptical challenges from anti-realists.} In this section, I will explain this effective approach to realism before turning to my challenges of it in Section 3 and 4.

Call the most radical realist view \textit{naive realism}, according to which our best scientific theories are literally true of our world, and in accepting it we should commit to every entity in the domain of the theories, which may include numbers, functions, manifolds, bundles, fields, particles, etc.  On the other end of the spectrum we have \textit{anti-realism} which says that the aim of science is only to provide successful empirical predictions, and we should not interpret theoretical statements in our scientific theories literally or believe in their literal truth.  Without going into any details (since the literature on this debate is humongous), I shall note that many people feel a strong pull towards some versions of scientific realism because it seems evident that science tells us a great deal more than (say) where the needle in a measuring device will turn and which part of a detecting screen will light up. It seems to be within the business of science to get the underlying stories about electrons, photons,  quacks and gluons behind those predictions at least approximately right. However, naive realism also seems highly implausible. As the famous pessimistic meta-induction goes,  even very successful and widely accepted theories in the past have been superceded by later theories. Thus there is no basis to believe that our current best theories will not be superseded by future ones. So it would be imprudent for us to commit to their exact truth. Besides,  scientific theories contain many uninterpreted statements that we cannot straightforwardly endorse realistically. For example, general relativity talks about tangent bundles, but it is not clear what aspects of reality these notions refer to (e.g., physical entities, platonic abstracta, or tropes).  
 
 In light of these problems, there is a strong motivation to find a middle ground. \textit{Selective realism} comes to the rescue, according to which only \textit{some} theoretical aspects of our best theories should be interpreted realistically. The idea is that we should believe only in those aspects of a theory that are unlikely to be rejected by future theories, those that are considered ``mature" and ``robust" (see Worrall 1989, Chakravartty 1998, Psillos 1999). For example, according to \textit{entity realism}, we should restrict ontological commitment to entities that have sufficient explanatory or causal power, while \textit{structural realism} says that we should only believe in certain relational structures between entities (though Chakravartty 1998 argues that these two amount to the same thing if plausibly construed). However, it has proven difficult to give a precise and systematic formulation of selective realism that does not run into major counterexamples and other troubles. For example, Stanford (2003) points out that many proposed criteria  for identifying the robust aspects of a theory are not prospectively applicable but only retrospectively so to past theories based on what aspects are actually retained (see also Laudan 1981).\footnote{For instance, some entity realists say that only terms with crucial explanatory power in a successful theory are generally referring, but what exactly is the criterion that Fresnel could use to correctly tell whether aether plays a crucial explanatory role in his relatively successful optical aether theory and should be retained? 
 	
 	As another example, Hacking (1982) argues that we should commit to an unobservable entity only if it can be manipulated to produce new experimental data. But the criterion in question is subsequently loosened to include entities that we do not interfere with or intervene on, like solar neutrinos, to rule out counterexamples.} There is little practical guide---barring hindsight---to discern the inventory of a theory worthy of ontological commitment.
 
Let's turn to the EFT approach and see what it implies for scientific realism. First, it is clear that realism towards EFTs is incompatible with naive realism. For one thing, the formulation of an EFT depends on regularization schemes, which are arbitrarily chosen. Interpreting an EFT realistically according to naive realism would entail that the chosen scheme  as well as all the quantities dependent on the scheme correspond to some physical reality, which is untenable. For example, as Williams points out,  in describing a free fermion field propagating in four-dimensional spacetime, we have to posit 16 ``mirror fermions'' in the effective Lagrangian if we place the theory on a spacetime lattice (i.e., impose a minimal length scale as the lattice spacing). These mirror fermions, which are products of a very specific scheme choice among many, are therefore not real but merely heuristic for yielding empirical predictions. Realism towards EFTs is also plainly incompatible with a version of realism that requires a scientific theory to be a \textit{complete} description of reality. EFTs have the built-in expectation of breaking down beyond a certain scale and therefore do not say anything about reality in that realm.

But these conflicts with naive realism turn out to be advantages if we implement selective realism instead. We can have more exact criteria for which entities in the domain of a theory as an EFT are robust and future-proof. First of all:
\begin{quote}
\textsc{Effective realism-1}. \textit{An entity (or a property, a relation) in the domain of a theory as an EFT is robust only if it is within the validity scope of the theory.}  
\end{quote}
The motivation behind this principle is clear: we cannot trust an EFT beyond its validity scope. Indeed, an EFT not only has a maximal energy level (or its equivalents) for its optimal applicability, but also a minimal energy level (or its equivalents).  Recall that according to the RG method, the physical couplings or parameters in an effective Lagrangian are optimized for a certain scale. To illustrate, an example discussed by Williams comes handy.  Quantum chromodynamics (QCD)---the QFT about the strong force---describes quarks and gluons (the force carrier) and their interactions at a relatively high energy level. At this level, their interactions are relatively weak, and an explanation in terms of quarks and gluons is appropriate. But at a lower energy level (corresponding to the length scale above $10^{-15}m$), the interaction between quarks and gluons becomes so strong that an explanation of experimental phenomena in terms of individuals states of quarks and gluons becomes intractable. Rather, the particles are \textit{confined} (a technical term referring to this kind of situation where interactions are very strong) into hadrons (particles that constitute nuclei such as neutrons and protons), and explanations in terms of hadrons become more appropriate. Thus, the picture of reality under this approach has multiple layers, each of which is described by an EFT of the relevant scale.\footnote{The corresponding picture of theories is called  ``tower of EFTs" (Cao and Schweber 1993):

\begin{quote}
	
	The necessity, as required by the decoupling theorem and EFT, of an empirical input into the theoretical ontologies applicable at the lower energy scales - scales to which the ontologies at the higher energy scales have no direct relevance in scientific investigations - is fostering a particular representation of the physical world. In this picture the latter can be considered as layered into quasi-autonomous domains, each layer having its own ontology and associated “fundamental” laws. (p.72)
	
\end{quote}
}

\textsc{Effective realism-1} offers a prospectively applicable necessary condition for ontological commitment that many other approaches lack. There are at least two benefits from adopting this principle in implementing selective realism. One is that an EFT with a validity scope is largely immune to future revisions because it is decoupled from unknown higher-energy theories. Regardless of what QFTs we will discover in the future at energy levels beyond our current access, or whether  the framework of QFT breaks down at those levels (e.g., being replaced by a version of string theory), it does not affect the validity of the QFTs we have for currently experimentally accessible levels. Second, the ontology within the validity scope generally plays a robust role in explaining relevant phenomena. Recall that in the example of confinement, what explains experimental data of a scattering process of hadrons involves hadrons rather than quarks and gluons that compose hadrons. So this principle satisfies the spirit of selective realism that we should only commit to entities that play important explanatory roles. 

Furthermore, not everything in the domain of validity should be endorsed realistically:

\begin{quote}
	\textsc{Effective realism-2}.  \textit{An entity (or a property, a relation) in the domain of a theory as an EFT is robust only if it is invariant under different regularization schemes or methods in general.}  
\end{quote}
The rationale behind this principle is straightforward. Mathematical artifacts resulting from our conventional choices like cut-off points should not be conflated with our physical reality. In Williams' example, mirror fermions and their behavior depend on our arbitrary choice of a regularization scheme and a  length cut-off, so we should not interpret them as physically real.   Like \textsc{effective realism-1}, this is a  prospectively applicable criterion for distinguishing between mathematical artifacts and physical entities in an EFT since every EFT is ``self-conscious'' about its regularization scheme and cut-off points. Note that making such a distinction is a general task of selective realism, since it is intuitive that some entities in a scientific theory (such as coordinate systems) are not worth ontological commitment independently from the concern of future revisions.

Call the view defined by \textsc{effective realism-1\&2} \textit{effective realism}. Are the two  criteria jointly sufficient for robustness? I will not go into this question, but merely note that a more stringent criteria for ontological commitment is certainly possible. For example, Fraser (2018) briefly suggests that only correlation functions are robust theoretical entities because all relevant observational data can be derived from them.

Now, the EFT approach to physics is based on the RG method and many scientific theories do not fit into the framework---for example, consider nonphysical theories like the theory of evolution, or many past physical theories like Newtonian mechanics. Two attitudes towards this are available. One is to generalize effective realism to all scientific theories considered as effective theories, namely equipped with a validity scope (and perhaps an analogous version of a regularization scheme).  The other, which is discussed  by Fraser (2018), is to treat selective realism on a case-by-case basis. The underlying  view is that there is no general strategy for implementing (selective) realism, so it is not a serious drawback that effective realism only focuses on a special group of theories considered in high energy physics. Both, I think, are reasonable approaches worth exploring, though I will not go further in this paper.  

To take stock, effective realism provides a concrete strategy for selective realism on how to interpret an EFT, and paints the picture of reality as consisting of different layers each boasting its own ontology.  Given empirically successful scientific theories treated as EFTs, we should commit to only those entities that are within their validity range and invariant under arbitrary choices of regularization schemes. This way, what makes EFTs ``effective'', namely that they deliberately ignore more fundamental degrees of freedom and include mathematical artifacts, is no longer a problem but becomes an essential part of this new approach to realism. As a robust doctrine of non-fundamental ontology, it may promise further benefits.\footnote{For example, it promises to give a new answer to the famous composition question, namely under what conditions material objects compose a further object. There are many attempts to answer the composition question, but a satisfactory answer is hard to come by, especially for the middle position that things compose but not always (see Korman and Carmichael 2016). The case of confinement gives some hint in that direction: there is a rigorous sense in which gluons and quarks compose hadrons. While it is impossible to work out all the details here, we can posit the following preliminary principle for composition in the framework of EFTs:
	\begin{quote}
		\textsc{Effective Composition.} For disjoint material objects $x$s, they compose a new object $y$ iff the interactions among $x$s are decoupled---through $y$---from all empirical predictions below a certain energy scale (or beyond a length scale). 
	\end{quote}
	Similarly as what motivates effective realism, this principle helps guarantee that the composite objects admitted into our ontology have sufficient or relatively basic explanatory, predictive, causal power with respect to a wide range of empirical phenomena. 
	
} However, I will now turn to a case study of applying this approach to an effective theory of spacetime, which proves to be problematic.

\section{Can We ``Effectivize'' Spacetime?}

The difficulty in unifying general relativity and quantum theory, as standardly construed, mainly lies in the (perturbative) non-renormalizability of gravity (or the spacetime metric). However, we do have a well-defined \textit{effective quantum gravity} (EQG) in the framework of EFTs. In light of this, many consider the difficulty surrounding quantum gravity to be much less drastic than it often sounds in the popular literature (see Burgess 2003, Donoghue 1995, 2012). It is clear that EQG does not give us a fundamental theory of spacetime. But under effective realism, can we at least extract a realistic picture of the spacetime metric from EQG at our currently experimentally accessible levels? I will argue not.

\paragraph{3.1 Effective quantum gravity (EQG)}

 The Einstein-Hilbert action for general relativity, from which Einstein's field equations are derived, can be approximately written as follows:
 $$\int d^{4}x\sqrt{g}(2/\kappa^{2})R$$
 where  $g$ is the determinant of the metric tensor $g_{\mu\nu}$, $\kappa$ is a coeffient with $\kappa^{2}=32\pi G$, and $R$ is the Ricci scalar curvature.\footnote{The full action is:
$$ S_{EH}=\int d^{4}x\sqrt{g}(\Lambda+2/\kappa^{2}R+c_{1}R^{2}+c_{2}R_{\mu\nu}R^{\mu\nu}+...+\mathcal{L}_{matter})$$
Here, $\Lambda$ is the cosmological constant, $R$ is the Ricci scalar curvature, $R_{\mu\nu}$ is the Riemann curvature tensor, and $\kappa, c_1, c_2$ are coeffients. The cosmological constant is relatively tiny, and so are $R^2, R_{\mu\nu}R^{\mu\nu},... $ such that these terms are typically irrelevant for physics (see Donoghue 2012). So the action is typically taken as $\int d^{4}x\sqrt{g}(2/\kappa^{2})R$, ignoring the matter fields.} This can be expanded into a perturbation series by taking the metric tensor $g_{\mu\nu}$ as the sum of a fixed background metric $\bar{g}_{\mu\nu}$ and a graviton field $h_{\mu\nu}$. $\bar{g}_{\mu\nu}$ is usually taken to be the Minkowski metric because for the scale of experimental setting where the scattering takes place, Minkowski spacetime is usually a good enough approximation for the background spacetime.\footnote{When a strong gravitational field is in question, then we need to work with a curved background spacetime. EFTs in curved spacetime are much less worked out and may be subject to difficulties depending on the properties of the curvature involved (see Burgess 2003)} Let $h$ be the canonically normalized graviton field $h=h_{\mu\nu}M_p$, where $M_p$ is the Planck mass. Then, expanding the Einstein-Hilbert action in terms of $h$ results in this:
$$S_{EH}=\int d^{4}x(\partial h)^{2}+1/M_{p}(\partial h)^{2}h+1/M_p^2(\partial h)^{2}h^2+...$$
According to Dyson's power-counting method, if a coupling constant entering the Feymann diagram has a negative dimension of mass, then the theory in question is non-renormalizable. In this case, the interaction terms have the coupling constant $1/M_p$ which has a negative dimension of mass. So the resulting theory is non-renormalizable. 

But this does not undermine the theory as an EFT. When the energy level $E$ that we use to probe the graviton field is far less than the Planck mass, then the non-renormalizable terms are suppressed by powers of $(E/M_p)$ and do not contribute much to the empirical results at the low energy level. In this case, we can still extract low-energy empirical predictions from the theory and match them with experimental data. For example, the amplitude of two gravitons scattering has been calculated up to two-loop approximation (Abreu et al 2020; for one loop approximation, see Dunbar and Norridge 1995).  It may be worth mentioning that the perturbative corrections to the leading terms are very small. So, as Donoghue (2012) argues, if small corrections are a positive feature of a QFT due to its resulting in more reliable calculations, then the quantum theory of gravity is the best QFT, not the worst! Like other EFTs, what happens at higher energy levels does not affect the scattering amplitude at the low energy level. So EQG seems as good as a renormalizable one within its validity range. Since we have independent reasons to endorse the EFT approach to physics, resorting to EQG is not an ad hoc move. 

\paragraph{3.2 Interpreting EQG}

 EQG just sketched is a partially successful quantization of gravity in the sense that it is a coherent mathematical machinery that yields empirical predictions within the experimentally accessible range. For this reason, many authors think that the crisis in physics pertaining to quantum gravity is overstated. On the other hand, despite the change of attitude towards EFTs in physics, many people still hold an instrumentalist attitude towards EQG, namely that the theory is merely a tool for yielding certain empirical predictions. For example, Crowther (2013) wrote that this approach ``sustains no illusions" and is ``just a means of combining GR and QM to make predictions in the regimes where we are able to.'' (p.327) It is certainly true that EQG, like other EFTs, is not a final theory. But an important question is left unaddressed: can we have a realistic reading of EQG through the lens of effective realism?

The quantization of the metric field that leads to EQG involves dividing it into a fixed background metric and a graviton fluctuation field:
$$g_{\mu\nu}= \bar{g}_{\mu\nu}+h_{\mu\nu}$$ As explained in 3.1, the background metric  $\bar{g}_{\mu\nu}$ is typically taken to be the Minkowski metric as a good approximation for the actual background metric where experiments take place. It is not yet completely clear in the physics literature what conditions the background metric needs to satisfy in order for the formalism of EQG to work, but for the sake of argument, I will grant that this formalism would work for  any sufficiently well-behaved background metric, including certain curved backgrounds.\footnote{Ordinary QFTs are only defined for fixed Minkowski spacetime, but can be expanded to some curved background (e.g., see Fewster and Verch 2015).}  The resulting theory is about the graviton field propagating in a fixed background spacetime. What realistic theory can we extract from it? Can we specify the ``basic'' ontology and laws of EQG that describe an intermediate level of reality? Let's consider the following options.

\paragraph{a. Standard interpretation}

An obvious candidate for being in the basic ontology of EQG is the graviton field $h_{\mu\nu}$. The reason for interpreting it realistically is that it is technically analogous to the particle fields in QFTs that we typically consider real. A graviton is an excitation of the metric field in the same formal sense as (say) a photon is an excitation of an electromagnetic field in quantum electrodynamics (QED). In particular, the graviton field would pass the robust test in effective realism, namely that the properties of the field do not depend on arbitrary choices of renormalization schemes and cut-off points.  Like the case of mirror fermions, the graviton field can be distinguished from its artificial counterparts. Recall that regulating QCD on a spacetime lattice would result in mirror fermions, the properties of which depend on the particular spacing of the lattice. As Williams explains, they can be separated from the real fermions: $$S_{lattice}=S_{continuum}+S_{mirror}$$ 
The total action $S_{lattice}$ formulated in terms of the lattice can be split into the action about fermions that do not depend on lattice and those that do. The mirror fermions are characterized by $S_{mirror}$, allowing us to interpret $S_{continuum}$ realistically. Similarly, gravitons can be distinguished from mirror gravitons in the lattice regulation and do not depend on choices of lattice spacing.

One may point out a disanalogy that in the case of matter theories there is usually no mention of the fixed background field unlike in the case of EQG. But this is because the background field is often the zero configuration that one omits mentioning. Analogously, the Minkowski metric may be considered the ``zero'' metric, except that the metric field cannot literally vanish anywhere unlike a matter field. However, it is possible to have a non-vanishing background matter field in an EFT (see Burgess 2003). So they are still quite analogous.

 Nevertheless, it seems problematic to consider the graviton field as part of the basic ontology for EQG. After all, it is a \textit{fluctuation} of the background field. If we take this claim literally, then the graviton field is ontologically dependent on the background field (since the fluctuation of a quantity is less fundamental than the quantity).  But why should we take this claim literally? Because the graviton field does not play any independent explanatory role in physical theories. It is the metric field as a whole that determines the spacetime geometry and symmetries of dynamic laws.  

Moreover, if we do admit the graviton field as a basic entity separately from the background metric, then it would lead to an unattractive dualistic reading of EQG: the background metric is naturally interpreted as representing fixed spacetime (fixed in the sense of not being subject to quantum effects rather than in the sense of not entering dynamic equations in classical general relativity) whereas the graviton field represents a quantum field propagating in spacetime. Such a dualism is typically rejected in interpreting general relativity, in which case we either consider a manifold  or the metric field (or both) as representing spacetime. (For example, Norton and Earman (1987) rejected the approach according to which spacetime is represented by the manifold equipped with Minkowski metric and the disturbance on the Minkowski metric represents a physical field.) For one thing, it would be less ontologically parsimonious by having two kinds of basic entities, namely the spacetime metric and the graviton field.\footnote{One may argue that the graviton field $h_{\mu\nu}$ is still part of spacetime because it is still part of the metric field $g_{\mu\nu}$ that we use to calculate distances in the larger scale. But the situation here is like the scenario of Poincare's disk where there is a background spacetime with its intrinsic flat metric and a universal force that distorts rigid rods and light rays such that spacetime \textit{appears} to be curved. Such a picture is more complicated than that of general relativity by its re-introduction of a universal force field eliminated by the latter. } At the very least, if we adopt this dualistic interpretation of EQG, we cannot say that EQG is a quantum theory of \textit{spacetime} since it is not spacetime or its metric that is treated quantum-theoretically.

Now, what if we read not the graviton field $h_{\mu\nu}$ but the whole metric field $g_{\mu\nu}$ realistically? The problem with it is that  $g_{\mu\nu}$ includes the background metric and thus makes reference to arbitrarily small scales. But the ontology of an EFT according to effective realism is restricted to its validity range. Recall that in the confinement example, gluons and quarks are part of the basic ontology of QCD which governs the physics at short distances within the experimentally accessible range. But at a sufficiently large length scale ($\approx 10^{-15}m$) at which their interactions get so strong that the individual particles become intractable, it is more appropriate to appeal to hadrons. Similarly, in EQG, we can admit only those entities whose sizes and interactions are at the appropriate length scale given by the validity range of the theory, which is well above the Planck level.

\paragraph{b. Truncating small distances}
Facing this problem, one solution is to only extract the metric field above a certain length scale appropriate for EQG into our ontology. The idea is that we keep only the distance relations between spacetime points determined by the metric field within the validity range. More precisely, let spacetime be modeled by $\langle \mathcal{M}, D\rangle$, where $\mathcal{M}$ is a classical manifold and $D$ is a metric that assigns a length to every path $c$ between manifold points in the usual way except that it has a extremal value given by the validity range of the theory: e.g., for spacelike paths, $D(c)=\int_{c}ds$  with $ds^2=\sum g_{\mu\nu}dx_\mu dx_\nu$, iff $\int_{c}ds\geq minimal$. The distance between two points is given by the length of the extremal path as usual.

But this strategy does not automatically get rid of arbitrarily fine-grained distances. For example, consider two points that do not have a distance relation under the above prescription, i.e., whose pre-interpretation distance is smaller than the minimal value. They can still have different distances to other points, and the difference could be a fragment of the minimal valid scale for EQG.  If we do not assign distances to ``overly close'' points, then why should we distinguish between such minuscule distance differences or between those points to begin with?

A perhaps more serious problem is that it is unclear if we can actually deal away with classical background metric defined at arbitrarily small scales. For example, when we calculate scattering amplitudes in EQG, we need to calculate the energy of the graviton field, which involves the classical background metric. More specifically, the Lagrangian for general relativity expanded in terms of the energy of the graviton field includes the following terms (Donoghue 1995, 10):\footnote{This is a more detailed expansion than the schematic expansion of $\mathcal{S}_{EH}$ mentioned in 3.1.}
$$\mathcal{L}_{grav}=\sqrt{\bar{g}}(...+\frac{1}{2}D_\alpha h_{\mu\nu}D^\alpha h^{\mu\nu}-\frac{1}{2}D_\alpha hD^\alpha h+...)$$ 
Here we can already see that the Lagrangian depends on the classical background metric in a number of ways. First, $D$ is a covariant derivative with respect to the background metric. Secondly, $D_\alpha h_{\mu\nu}D^\alpha h^{\mu\nu}$ is the inner product of $D_\alpha h_{\mu\nu}$ with itself, which depends on the background metric (the metric is used to raise the index of $D_\alpha h_{\mu\nu}$ before it can be contracted with the original). Similarly, $D_\alpha hD^\alpha h$ and even $h$ alone (which is the contraction of the graviton field with itself) depend on the background metric. There is also the explicit mention of the classical volume form $\sqrt{\bar{g}}$. Thus, if we truncate the classical metric field in the suggested way, then we won't be able to carry out these calculations needed in EQG. Thus, truncating the classical field has undesirable ramifications for the formalism of EQG.

Furthermore, the RG method also involves a classical background metric (though not necessarily the same one as in EQG). In Section 1, I explained how physical coupling constants depend on the energy or length scales, but the scales are determined by the background metric. Insofar as the appeal of EQG is dependent on this framework, EQG is indirectly dependent on the background metric of RG. Thus, even if we can truncate $g_{\mu\nu}$ in EQG in the suggested way, we haven't eliminated altogether the theory's dependence on a classical metric defined at arbitrarily small distances. While for a realistic reading of EQG advocated by the effective realism, we do not need to register all the things it quantifies over,  there is something particularly unsatisfying in that when we extract a theory of metric from EQG, we are unable to interpret the classical background metric that does robust work in the theory.

It may be possible to purge EQG---and in general the EFT framework---of all references to classical background metrics by developing a new framework of EFTs on a spacetime lattice. Instead of truncating small distances, we start with a lattice and define all quantum fields including the metric field on it.\footnote{For example, lattice QFT (LQFT) is such a formalism (see Montvay and Münster 1997). Note that, however, the lattice involved in LQFT is typically not intended to be read realistically (e.g., lacking important symmetries and having unusual signature---Euclidean rather than Lorentzian---or fewer dimensions), but as a mere computational device.} But this attempt is better described as developing an alternative framework rather than an interpretation of EQG. Indeed this is part of developing a full theory of quantum gravity (for example, loop quantum gravity is in its vicinity; see Rovelli and Viditto 2015). So it seems that we are running out of options for extracting a realistic ontology of  spacetime metrics from EQG. 

There are two possible implications of this conclusion. First, perhaps effective realism is a good approach for other QFTs as EFTs but not for EQG, which would indicate that EQG is defective. This would be an interesting result in that it helps pin down what is particularly wrong with EQG comparing to other EFTs and helps counter the complacency towards EQG.  Second, perhaps the defect lies not in EQG but in effective realism (or both). I am inclined to think that the problematic case of EQG indeed undermines effective realism to some extent. Moreover, it also brings out a general problem for interpreting any EFTs, which I will turn to now. 
 
\section{A General Challenge to Effective Realism}

In this section, I will examine whether the problems in applying effective realism to EQG raise general issues for this approach. 

 Let's reflect on the relevant similarities and dissimilarities between EQG and other EFTs in particle physics, especially QFTs. As EFTs, all these theories are very much on a par. To start with, other EFTs also have the analogous problem of classical background \textit{matter} field. More detailedly, the quantization of a classical matter field in QFTs also involves dividing it into a fixed background field and its fluctuation:
 
 $$\phi=\bar{\phi}+\psi$$
 Like in EQG, the background field $\bar{\phi}$ is usually taken to be the classical vacuum state of the field. This can be omitted and thereby does not lead to an interpretative issue  (the vacuum state of a matter field vanishes everywhere while that of a metric field is the Minkowski metric). However, those background fields sometimes are not vacuum or zero states (QFTs with non-zero background fields are not commonly considered due to calculation complexity, but entirely possible). In this case, we need to interpret this fixed background matter field. However, this challenge can be met more easily than in the case of EQG because the references to classical background matter fields in arbitrarily small regions can be truncated---unlike classical metrics, classical matter fields are not part of the framework for EFTs.\footnote{The truncation can be implemented by, for example, discarding all Fourier components at momenta larger than a cut-off.}   
 
What's more troublesome, like EQG, other EFTs also involve a classical background \textit{metric} field, which is likewise involved in calculating the energy of matter fields. For example, we may similarly interpret QFTs as describing certain particles propagating in classical Minkowski spacetime (or some suitably curved spacetime), which is defined to arbitrary accuracy. How should we interpret this reference? First, it might be worth noting that the problem here is not exactly analogous with the case of EQG. To remind: interpreting EQG as gravitons propagating in classical spacetime commits an unpleasant dualism between the graviton field and spacetime metric, which we standardly accept to be one and the same entity.  In other EFTs, there is no such unwanted dualism. However, we still face a dilemma over whether we should take the background spacetime realistically. On the one hand, classical spacetime does not satisfy the condition for ontological commitment that it fall under the validity range of the EFTs, since arbitrarily small regions are parts of classical spacetime. On the other hand, it also does not satisfy the criterion for a non-entity because it does not vary under different choices of cutoff points or regularization schemes. So it seems that classical spacetime is neither a physical entity nor a mathematical artifact by the standards that we have discussed. So what should we think of it? 
 
 One move is to consider classical spacetime as part of the heuristic device to get the empirical predictions rather than carrying genuine ontological weight in EFTs. The idea is that \textsc{effective realim-1\&2} are necessary conditions but not jointly sufficient for ontological endorsement, and therefore we are not obligated to endorse spacetime under effective realism. This may seem reasonable at first. As quantum theories of matter, their failure to include a realistic ontology of spacetime is no failure as realistic theories of matter, since it is unreasonable to expect a scientific theory to be an exhaustive description of reality. Nevertheless, this strategy is unsatisfactory. An interpretation of a scientific theory should allow it to ``discharge all its scientific duties.''(Ruetsche 2011) Clearly, one of the duties that QFTs have is to explain the observable phenomena in terms of the particle fields and their dynamics that are probeable at currently experimentally accessible scales. The background spacetime plays an important role in this explanation. For instance, if we omit spacetime from the explanation for certain electromagnetic phenomenon such as ``a ray of electrons with this-and-that properties propagating in Minkowski spacetime in such-and-such a way,'' the explanation would not be complete.

 I would like to emphasize the distinction between this problem and one that Fraser (2009) raises and Williams (2019) rejects. Fraser argues that to interpret the lattice formulation of EFTs  realistically, we are obliged to take spacetime to have this lattice structure. But since we should not believe that spacetime has this lattice structure, we should refrain from interpreting those EFTs realistically. This reasoning relies on the rejected assumption that a theory needs to describe reality exhaustively: since an EFT does not say anything below a certain length scale, this means that there is a minimal length. Williams rightfully rejects this assumption in favor of effective realism, according to which a theory only needs to describe reality within its validity scope. Besides, even if there is a minimal length, it does not immediately follow that spacetime is composed of lattice points, as discussed in 3.2(b). In contrast, the problem I raise here is about the interpretative problem of background spacetime, not of the lattice structure. 
 
 To take stock, we face a dilemma in interpreting the reference to classical background spacetime in any EFT: it should not be included in the ontology because it is defined beyond the validity scope; but it should be included in the ontology because it plays a necessary explanatory role. For matter fields, the references to their high-energy (or short-length) parts are dispensable---which is part and parcel of the EFT approach. But for spacetime, the references to smaller regions are indispensable, at least in the current framework of EFTs.

\section{Conclusion}

Do we have a quantum theory of spacetime metrics at the experimentally accessible small scales where quantum effects are prominent?  I have argued for ``no,'' despite the fact that effective quantum gravity (EQG) is celebrated as a satisfactory physical theory at these scales. A natural interpretation of EQG says that the graviton field, which is subject to quantum fluctuations, propagates in classical background Minkowski spacetime---which is not a quantum theory of spacetime metrics. This also raises a general concern for effective realism, the view that ontological commitment of a theory is restricted to its validity scope, which is otherwise attractive and even imperative given the EFT approach to physics. The classical background spacetime involved in EFTs neither is a clear non-entity nor belongs to the layered ontology licensed by those EFTs.


\newpage
	
	\begin{thebibliography}{35}

\bibitem{Ab20} Abreu, S., F. Febres Cordero, H. Ita, M. Jaquier, B. Page, M. S. Ruf, V. Sotnikov. (2020) ``The Two-Loop Four-Graviton Scattering Amplitudes."  	arXiv:2002.12374 [hep-th].


\bibitem{Br11} Braun, J., H. Gies, and D. Scherer. (2011) ``Asymptotic safety: a simple example'' \emph{Physical Review D}, 83(8).

\bibitem{Bu03} Burgess, C.P. (2003). “Quantum Gravity in Everyday Life: General Relativity as an Effective Field Theory,” \emph{Living Reviews of Relativity}.

\bibitem{Bu14} Butterfield, J. and Bouatta, N. (2014). “Renormalization for Philosophers,” \emph{Metaphysics in Contemporary Physics}. 


\bibitem{Ca93} Cao, T.Y., Schweber, S.S., (1993) ``The conceptual foundations and the philosophical aspects of renormalization theory,'' Synthese 97 (1):33 - 108.




\bibitem{Ch97} Chakravartty, A. (1998) ``Semirealism,'' Stud. Hist. Phil. Sci., Vol. 29, No. 3, pp. 391–408.



\bibitem{Cr13} Crowther, K. (2013) “Emergent spacetime according to effective field theory: From Top-down and Bottom-up” \emph{Studies in History and Philosophy of Modern Physics Part B} 44(3). 321-328.

\bibitem{Cr16} Crowther, K. (2016) \emph{Effective Spacetime: Understanding Emergence in Effective Field Theory and Quantum Gravity}. Springer. 

\bibitem{Cr17} Crowther, K. and Linnemann, N. (2017). “Renormalizability, Fundamentality, a Final Theory: the role of UV-completion in the search for quantum gravity,” \textit{British Journal for Philosophy of Science}.

\bibitem{Co11} Costello, K. (2011). \emph{Renormalization and Effective Field Theory}. American Mathematical Society.



\bibitem{Do95} Donoghue, J. (1995). ``Introduction to the Effective Field Theory Description of Gravity.'' arXiv:gr-qc/9512024.
		
\bibitem{Do12} Donoghue, J. (2012). “The effective field theory treatment of quantum gravity.”  arXiv:1209.3511 [gr-qc].

\bibitem{Du95} Dunbar, D.C. and Norridge, P.S., (1995) “Calculation of graviton scattering amplitudes using string based methods,” Nucl. Phys. B 433, 181 [hep-th/9408014].

\bibitem{Du12} Duncan, A. (2012): \emph{The Conceptual Framework of Quantum Field Theory}. Oxford: Oxford University Press.


\bibitem{en87} Earman, J., Norton J. (1987). “What price spacetime substantivalism? The hole story”. \emph{British J. Philos. Sci.} 38.4, pp. 515–525.

\bibitem{Fe15} Fewster, Christopher J. and Verch, Rainer (2015) ``Algebraic quantum field theory in curved spacetimes.'' arxiv.1504.00586.

\bibitem{Fr09} Fraser, D. (2009) ``Quantum Field Theory: Underdetermination, Inconsistency, and Idealization,'' \emph{Philosophy of Science}, 76, pp. 536–67.

\bibitem{Fr18} Fraser, J. (2018) ``Renormalization and the Formulation of Scientific Realism'' \emph{Philosophy of Science}, 85 (5),  pp. 1164 - 1175.

\bibitem{Ha82} Hacking, I. (1982) ``Experimentation and Scientific Realism,'' \emph{Philosophical Topics}, Vol. 13, No. 1, Realism, pp. 71-87.

\bibitem{He15} Hertel, I.V., Schulz, CP. (2015). ``Basics.'' In: \emph{Atoms, Molecules and Optical Physics 1.} Springer, Berlin, Heidelberg.






\bibitem{Ko16} Korman and Carmichael (2016) ``Composition,'' \emph{Oxford Handbooks Online}.

\bibitem{La80} Laudan, L. (1981) ``A Confutation of Convergent Realism,'' \emph{Philosophy of Science}, Vol. 48, No. 1, pp. 19-49.

\bibitem{Ms21} Marzocca, D.,  Trifinopoulos, S. (2021) ``Minimal Explanation of Flavor Anomalies: B-Meson Decays, Muon Magnetic Moment, and the Cabibbo Angle,'' \emph{Phys. Rev. Lett.} 127, 061803.

\bibitem{Mo97} Montvay I. and G. Münster (1997) \emph{Quantum Fields on a Lattice}, Cambridge University Press


\bibitem{Pe95} Peskin and Schroeder (1995) \emph{Introduction to Quantum Field theory.} Perseus Book Publishing.  


\bibitem{Pe18}  Petreczky, P., Taku Izubuchi, Luchang Jin, Christos Kallidonis, Nikhil Karthik, Swagato Mukherjee, Charles Shugert, Sergey Syritsyn (2018) ``Pion structure from Lattice QCD,'' arXiv:1812.04334 [hep-lat].

\bibitem{Ps99} Psillos, S. (1999) \emph{Scientific Realism: How Science Tracks Truth}. London: Routledge.

\bibitem{Ru18}  Ruetsche, L. (2018), ``Renormalization Group Realism: The Ascent of Pessimism," \emph{Philosophy of Science}. 85:5, 1176-1189. 

\bibitem{Ru11} Ruetsche, L. (2011) \emph{Interpreting Quantum Theories}, Oxford: Oxford University Press.

\bibitem{Rv15} Rovelli and Vidotto (2015) \emph{Covariant Loop Quantum Gravity: an elementary introoduction to quantum gravity and spinfoam theory}. Cambridge. 


\bibitem{St03} Stanford, K. (2003), ``Pyrrhic Victories for Scientific Realism,'' \emph{Journal of Philosophy}, 100(11): 553–572.


\bibitem{We95} Weinberg, S. (1995), \emph{The Quantum Theory of Fields, volume 1}. Cambridge University Press.

\bibitem{We99} Weinberg, S. (1999), ``What is quantum field theory and what did we think it was,'' in Cao ed., pp. 241-251. Also at: arxiv: hep-th/9702027.

\bibitem{Wi79} Wilson, K. (1979), ‘Problems in Physics with Many Scales of Length’, \emph{Scientific American} 241: 158-179.

\bibitem{Wi17} Williams, P. (2019) ``Scientific Realism Made Effective.'' \emph{British J. Philos. Sci.} 70 (1): 209-237.

\bibitem{Wi18} Williams, P. (2021) ``Renormalization Group Methods.''  \emph{Routledge Companion to Philosophy of Physics}, Eleanor Knox and Alastair Wilson (eds). 

\bibitem{Wo89} Worrall, John, (1989) ``Structural Realism: The Best of Both Worlds?,'' \emph{Dialectica}, 43(1–2): 99–124. 



		
		
	\end{thebibliography}
\end{document}